\documentclass[preprint2]{aastex}

\singlespace
\slugcomment{slug comment}
\shorttitle{short title}
\shortauthors{short authors}

\begin{document}

\title{Relativistic Outflow of Electron-Positron Pair Plasma 
from a Wien Equilibrium State}
\author{S. Iwamoto and F. Takahara}
\affil{Department of Earth and Space Science, Graduate School of Science, Osaka University}
\affil{Machikaneyama 1-1, Toyonaka, Osaka, 560-0043, Japan}
\email{iwamoto@vega.ess.sci.osaka-u.ac.jp}
\email{takahara@vega.ess.sci.osaka-u.ac.jp}

\begin{abstract}

We investigate a novel mechanism of bulk acceleration of relativistic 
outflows of pure electron positron pairs both analytically and numerically.  
The steady and spherically symmetric flow is assumed to start from a Wien 
equilibrium state between pure pairs and photons in a compact region 
which is optically thick to the electron scattering at a relativistic 
temperature. Inside the photosphere where the optical thickness becomes 
unity, pairs and photons behave as a single fluid and are thermally 
accelerated. Outside the photosphere, pairs and photons behave separately 
and we assume the free streaming approximation for photons, which are 
emitted from the relativistically moving photosphere. Pairs are shown to 
be thermally accelerated further even outside the photosphere because 
the photospheric temperature is at least mildly relativistic. It is to 
be noted that the mean energy of photons is higher than that of pairs 
in the comoving frame of pairs and that the Compton interaction leads to 
additional heating and radiative acceleration of pairs. For a reasonable 
range of the boundary temperature and optical thickness, the terminal 
Lorentz factor of pair outflows turns out be more than 10 and the terminal 
kinetic power accounts for more than 2/3 of the total luminosity. 
While the total luminosity should be at least larger than the Eddington 
luminosity, real luminosity can be modest if the outflow is collimated by 
some unknown mechanisms. This mechanism successfully avoids the 
difficulties of pair annihilation and radiation drag owing to the pair 
production by accompanying high energy photons and the strong beaming of 
radiation field. It is seen that most pairs injected at the boundary 
survive to infinity. The radiation from the photosphere should be 
observed as MeV peaked emission at infinity with an order of kinetic 
power of jets.

\end{abstract}


\notetoeditor{note to editor}

\section{Introduction}

The production and bulk acceleration of relativistic jets in active 
galactic nuclei is one of the most challenging problems in astrophysics. 
The basic properties to be explained are the highly relativistic velocity 
with the bulk Lorentz factor of around 10, the kinetic power almost 
comparable to the Eddington luminosity, and the collimation into a small 
opening angle \citep{OSMG,BBR84}. 
Although many ideas have been proposed ranging from magnetic and radiative 
accelerations to  
Blandford-Znajek process, each model has its own difficulties 
and there is no general consensus on how jets are produced and accelerated 
\citep{B95,BBR84}. In this paper, we explore a thermal acceleration 
mechanism of pure electron-positron plasma in view of the recent 
observational and theoretical developments. 

Several important observational developments have been made in this decade. 
First of all, EGRET on board the Compton Gamma Ray Observatory 
has found that blazars are a strong $\gamma$-ray emitter \citep{M97}.
The $\gamma$-ray emission is interpreted in terms of the inverse 
Compton scattering off relativistic electrons in the jets the direction 
of which is so close to the line of sight that the emission from jets 
is strongly beamed \citep{blan79,MGC92,SBR94}. 
Combining the synchrotron emission at lower energies, 
we can estimate the physical states of the emission region in the jets. 
The results of multi-frequency spectral fitting indicate that the jets are 
particle dominated rather than magnetic field dominated \citep{IT96,KTK01}. 
The energy density of relativistic electrons is more than one order of 
magnitude larger than that of the magnetic field. The size of the emission 
region is estimated to be $10^{16} \sim 10^{17} {\rm cm}$ 
from time variabilities and visibility of $\gamma$-ray photons. 
Relativistic beaming factor of around 10 is required to assure 
the visibility of $\gamma$-ray photons. Since the distance of the emission 
region from the central black hole is reasonably estimated to be 
10 times its size, it is around $10^{17} \sim 10^{18} {\rm cm}$, 
an order of magnitude shorter than that of superluminal knots 
observed at radio wavelengths.   
Thus, the bulk acceleration of jets must occur within such short distances. 
As for the matter content of jets, electron-positron pair dominance has been 
proposed based on several independent observations \citep{W99,reyn96,H99,H01}, 
although the contribution of protons is not completely excluded.   

Second important development is the estimation of the kinetic power of 
extended radio sources. Such studies have shown that large scale kinetic 
power is well correlated with radiative power and suggest that jets are 
produced by accretion process rather than they are two unrelated phenomena 
\citep{RS91}. 
Related to this, Galactic superluminal sources GRO J1655+40 and 
GRS 1915+105 with the bulk Lorentz factor of a few were discovered 
\citep{hjel95,mira94}. The jet ejection for these two sources seems to 
be correlated with X-ray behavior, which suggests that the jets are 
really produced by accretion \citep{FMP99}.

Finally, developments of observations and models of cosmic gamma-ray 
bursts (GRBs) are quite fascinating. Confirmation of the afterglows of 
GRBs implies the basic correctness of the fireball model 
\citep{pira93, RM92}. 
In this model, GRBs are initiated by the production of 
a compact high entropy fireball, although no specific mechanism of 
the production process itself has been widely accepted.
The fireball is relativistic thermal plasma, which consists of 
photons, electron-positron pairs and a small amount of baryons. 
As the fireball expands, it is thermally accelerated and 
the temperature drops. When the temperature drops below the electron 
mass, electron-positron pairs annihilate, but it is  
still optically thick to the electron scattering because of a 
small baryon load. Final outcome is that the initial thermal energy 
is converted to the bulk kinetic energy of baryons attaining the bulk 
Lorentz factor of $E/M_{\rm b}c^2$, where $E$ and $M_{\rm b}$ are the 
total energy and baryon mass of the fireball. 
Inhomogeneities in the fireball are expected to produce internal shocks 
which are supposed to correspond to GRBs themselves and 
an impact to ambient matter produces an external shock which 
corresponds to afterglows. The emission from shocks is ascribed to 
the particle acceleration by shocks. 
It is also mentioned that some GRBs 
are described by jets rather than spherical explosion with respect 
to the afterglow light curve and energetics.

We believe that the success of the fireball model for GRBs 
has strong implications for jets in blazars. 
For blazars, internal shocks correspond to the compact emission 
regions which are located relatively near the central black hole and 
external shocks produce large-scale radio sources \citep{CB96, BC89}. 
But, one strong difference between blazars and GRBs is the size of 
the fireball. The initial size of GRB fireballs is $10^6 \sim 10^7$ cm 
and it is in a complete thermal equilibrium for the total 
energy of $\sim 10^{52}$erg. In contrast, the size of blazar fireballs 
should be around 3$r_{\rm g}\sim 10^{14}$cm with a similar total energy, 
where $r_{\rm g}\equiv 2GM/c^2$ is the Schwarzshild radius for 
the black hole mass $M$; the typical value of $M$ is taken to be 
$10^8 M_\odot$. 
If the blazar fireball is in a complete equilibrium, the temperature 
is an order of $10^5$K which is far below the electron mass and 
no pair production is expected. 
Basically it describes an optical thick 
radiation pressure dominated standard accretion disk and 
no relativistic expansion is expected. 

However, for an optically thin state, electron temperature can be 
relativistic and copious electron-positron pairs 
can exist in principle. Even for mildly relativistic temperatures, 
copious electron-positron pairs exist for a finite optical thickness 
to scattering \citep{sven84,ligh82,KT85}. It has been shown that 
the pair density is limited to a relatively small value in hot accretion disks 
when the pair production is balanced with pair annihilation \citep{KT88}. 
However, effect of pair escape can increase the pair density as was exemplified 
by \cite{yama99}. They have argued that pair outflow from hot accretion disks 
may account for major part of the dissipated energy from accretion. 
In such a state, photons and electron-positron pairs are not in a complete 
thermal equilibrium but they are coupled by electron scattering 
and pair production/annihilation process and the photon density is far below 
that of the equilibrium state for a given temperature. 
Thus, the escaping pairs just above the disk may act as the fireball 
to produce relativistic jets.  
Although the physical state of such escaping pairs has not been fully 
investigated by now, we believe that this is the most promising way 
of producing relativistic jets with electron-positron pairs. 

In this paper, we consider the bulk acceleration of such fireballs 
starting from a range of the initial states in the simplest way. 
We treat steady and spherically symmetric outflows starting from 
pure electron-positron pairs with photons in a Wien equilibrium 
for which photon distribution function is given by a Boltzmann 
distribution with a finite chemical potential.  
The formulation is similar to that of \citet{grim98} (GW98), 
in which relativistic 
flows of pure electron-positron pairs 
starting from a complete thermal equilibrium were analyzed.
The dynamics starting from a Wien equilibrium is expected to be similar to 
that given by GW98 when proper accounts for the difference are taken. 
This type of flows takes into account of the effects of pair production 
as well as pair annihilation. Then, we can examine to what extent we 
can avoid the pair annihilation problem pointed out by some authors 
arguing against electron-positron pair dominance in relativistic jets 
\citep{CF93,BL95}. 
If pairs eventually annihilate in the course of expansion, there would 
remain very little kinetic power. It is very important to quantitatively 
assess the surviving kinetic power of such a fireball.

In section 2, we give basic formulation, in section 3 we present approximate analytical solutions and in section 4 we present the 
results of numerical solutions.

\section{Formulation}

We assume that at the inner boundary the source of pure electron-positron 
pairs and high energy photons injects a steady outflow of pairs 
and radiation. The optical thickness of such a region to the electron 
scattering is large but that to the absorption is smaller than unity. 
Thus, the physical state of such a regime may be characterized by 
a Wien equilibrium in the simplest case. When we neglect the dynamical 
effects, physical state of such a plasma with a finite amount of protons 
is known to be strongly constrained by thermal balance and pair equilibrium 
owing to the inefficiency of pair annihilation at  
relativistic temperatures \citep{ligh82,sven84,KT85}. 
However, we should note that the time scale to 
reach such a steady state is fairly long. 
If the plasma is first put at a relativistic temperature without pairs, 
pair production proceeds very quickly and the subsequent photon production 
and Compton scattering strongly cool the plasma  
as the pair concentration still increases. 
Finally, pair annihilation leads to a pair equilibrium state. 
The time scale to reach the pair equilibrium state is controlled by the photon 
production and photon escape, the time scales of which are much longer 
than the dynamical time scale. 
Thus, the dynamical effects are essential to the behavior of such plasma 
and most probably lead to the expansion and outflow. 
The constraints on the electron temperature and optical thickness 
should be drastically changed from those for static confined plasma. 
In this paper, for simplicity, we assume that a Wien equilibrium state is 
already established from the outset and that the temperature and 
optical thickness can take a wider range than those of static cases. 
Since the properties of the initial states, where dynamical effects are 
critically important, are not well known, we here regard the temperature 
and optical thickness at the boundary as free parameters. 

The flow is optically thick near the boundary and becomes optically thin 
beyond the photosphere. We treat separately optically thick and thin regimes 
as GW98.
For optically thick regime in which the optical thickness to 
the electron scattering for radially outgoing photons is larger than unity, 
we treat as if electron-positron pairs and photons form a single fluid.
Although, radiation field may be anisotropic, we simply assume isotropic photon distribution in the comoving frame.
Note that as was explicitly shown in GW98, the isotropy of photon field is a surprisingly good assumption in a relativistic expanding flow.
Although the assumption of a Wien equilibrium is of course a simplification
to treat the dynamical problem in a simple way, we may regard this a reasonable
choice for a relativitic expanding flow if the flow start from a Wien equilirium.
This is seen in the behavior of the cosmic microwave background radiation in
cosmology and the optically thick relativistic wind considered by GW98;
the photon distribution remains a Planckian even after the optical thickness
becomes much below unity. Anyhow, to obtain more exact photon spectrum,
we need to solve the radiative transfer coupled with dynamics, a challenging
problem which will be pursued in future.
Then, the basic equations are the conservation laws of energy, momentum 
and numbers.
For a steady and spherically symmetric flow, conservation laws of 
energy and momentum are given by
\begin{equation}
{1 \over r^2}{d \over dr}
\left[ r^2 (\rho+P) \Gamma^2 \beta \right]=0
\label{2-1}
\end{equation}
and
\begin{equation}
{1 \over r^2}{d \over dr} 
\left[ r^2 (\rho+P) \Gamma^2 \beta^2 \right] 
+{dP \over dr} =0,
\label{2-2}
\end{equation}
respectively. Here, $\rho$ and $P$ are the total energy density and pressure 
of electron-positron pairs and photons.
Both components are described by ideal fluids and we use the units of $c=1$.
The bulk flow velocity and its
Lorentz factor are denoted by $\beta$ and $\Gamma$, respectively.
Momentum conservation law is alternatively written by 
\begin{equation}
\frac{1}{\Gamma}\frac{d\Gamma}{dr}+\frac{1}{\rho+P}\frac{dP}{dr}=0.
\label{2-3}
\end{equation}

The energy density and pressure of electron-positron pairs are given by
\begin{equation}
\rho_{\rm e \pm}=2 m_{\rm e} n_{\rm e} \langle \gamma \rangle
\label{2-4}
\end{equation}
and
\begin{equation}
P_{\rm e \pm}=2 m_{\rm e} n_{\rm e} \theta,
\label{2-5}
\end{equation}
respectively. Here, $n_{\rm e}$ is the number density of electrons 
(=that of positrons) and the total number density of leptons are 
$2n_{\rm e}$, 
$\theta$ is the temperature normalized by the electron mass $m_{\rm e}$ and 
$\langle \gamma \rangle=K_3(1/\theta)/K_2(1/\theta)-\theta$ is 
the average Lorentz factor of electron thermal velocity, 
where $K_i$ is the $i$-th order modified Bessel function of the second kind.

As was mentioned above, we assume that pairs and photons 
are in a Wien equilibrium in the initial state. 
In a Wien equilibrium, the energy density and pressure of photons are 
given by
\begin{equation}
\rho_\gamma=3 m_{\rm e} n_\gamma \theta
\label{2-6}
\end{equation}
and
\begin{equation}
P_\gamma= m_{\rm e} n_\gamma \theta,
\label{2-7}
\end{equation}
respectively, where $n_\gamma$ is the number density of photons.
In a Wien equilibrium, $n_{\rm e}$ and $n_\gamma$ are related by
\begin{equation}
{n_e \over n_\gamma} = {K_2 (1/\theta) \over 2 \theta^2} \equiv f(\theta).  \label{suumitudohi}
\end{equation}

In numerical calculations of the flow we do not use eq.(\ref{suumitudohi}), but
solve the following number conservation equations, 
assuming that pairs and photons take the same temperature,
\begin{equation}
{1 \over r^2}{d \over dr}
\left[ r^2 n_e \Gamma \beta \right] = \dot{n}_{\rm e} 
\label{2-9}
\end{equation}
and
\begin{equation}
{1 \over r^2}{d \over dr} 
\left[ r^2 n_\gamma \Gamma \beta \right] = \dot{n}_\gamma,
\label{2-10}
\end{equation}
where $\dot{n}_{\rm e}$ is the pair creation rate minus annihilation rate.
The pair annihilation rate is approximated by 
\begin{eqnarray}
\left( \dot{n}_{\rm e} \right)_{\rm ann} &=& 
\langle \sigma_{\rm ann} v \rangle n_{\rm e}^2  \\
&=&{3 \over 8} \sigma_T {n_{\rm e}}^2 
\left[ 1+ {2 \theta^2 \over \ln(1.12 \theta + 1.3)} \right]^{-1}, \nonumber
\end{eqnarray}
where $v$ denotes the thermal velocity of pairs \citep{sven82}. 
The pair creation rate is given by 
\begin{eqnarray}
\left( \dot{n}_{\rm e} \right)_{\rm cre}  &=&  
\left( {n_\gamma \over n_{\rm e}} \right)^2 
\left[ {K_2(1/\theta) \over 2 \theta^2} \right]^2 
\left( \dot{n}_{\rm e} \right)_{\rm ann},
\end{eqnarray}
when the distribution function of photons is a Wien one \citep{sven84}. 
Since we neglect other photon production processes such as bremsstrahlung, 
there is a simple relation $2\dot{n}_{\rm e}=-\dot{n}_\gamma$.

Then, we can solve for the bulk velocity, temperature and electron 
number density from the energy, momentum and number conservation 
equations given above.
As the boundary conditions, we give the bulk velocity, temperature, and electron 
number density at the inner radius at $r=r_0=2 r_g$. 
The optical thick approximation is valid between the inner 
radius and the photosphere.
The radial coordinate of the photosphere $r_{\rm ph}$ is defined by
\begin{equation}
\tau= \int^\infty_{r_{\rm ph}} dr \ 2n_e \sigma_T \Gamma (1-\beta) =1.
\label{tau}
\end{equation}
The optical thickness depends on the direction of the photon and 
for photons moving oblique to the radial direction the optical 
thickness is larger than that given above.
But within the cone of $1/\Gamma$ around the radial direction 
the difference is small.
Outside the cone, the optical thickness becomes much larger.
Since we are treating relativistic outflow, the most relevant 
is the radial optical thickness.

In the optically thin region we treat pairs and photons separately.
For pairs, energy, momentum and number conservation laws are given by
\begin{eqnarray}
 {1 \over r^2}{d \over dr} 
\left[ r^2 (\rho_{\rm e \pm} + P_{\rm e \pm}) \Gamma^2 \beta \right] 
= F^0, && \label{2-14}  \\
{1 \over r^2}{d \over dr} 
\left[ r^2 (\rho_{\rm e \pm} + P_{\rm e \pm}) \Gamma^2 \beta^2 \right] 
+ \frac{dP_{\rm e \pm}}{dr} = F^1 && \label{2-15}
\end{eqnarray}
and
\begin{eqnarray}
 {1 \over r^2} {d \over dr} 
\left[ r^2 n_{\rm e} \Gamma \beta \right] 
&=& \dot{n}_{\rm e}. \label{2-16}
\end{eqnarray}
The right-hand side of energy and momentum conservation laws denotes 
the so-called radiative force, which is given through 
the Lorentz transformation of the radiative force in the fluid frame. 
The latter for Compton cooling is given by \citet{phin82} and \citet{siko96}
\begin{equation}
({F'^{0}})_{\rm rad} = -{4 \over 3} (2n_{\rm e}) \sigma_T 
(\langle \gamma^2-1 \rangle - \epsilon_\gamma) {T'^{00}_\gamma}
\end{equation}
and
\begin{equation}
({F'^{1}})_{\rm rad} = (2n_{\rm e}) \sigma_T 
\left\{ {1 \over 3}+{2 \over 3} \langle \gamma^2 \rangle \right\} 
{T'^{01}_\gamma},
\end{equation}
where $\epsilon_\gamma$ denotes the phenomenological Compton heating term  
described below and $T_\gamma^{\mu\nu}$ is the energy momentum tensor 
of radiation field. 
The prime denotes the quantities in the fluid frame 
whereas those of 
the central black hole frame are represented without prime. 
Lorentz transformation between these frames is described as 
\begin{equation}
F^\mu={\Lambda^\mu}_{\nu} F'^{\nu} \ {\rm and} \
T^{\mu\nu}={\Lambda^{\mu}}_{\delta} {\Lambda^{\nu}}_{\sigma} 
T'^{\delta \sigma},
\end{equation}
where
\begin{equation}
{\Lambda^\mu}_{\nu}=
\left(
\begin{array}{cc}
\Gamma & \Gamma\beta \\
\Gamma\beta & \Gamma
\end{array}
\right).
\end{equation}

The Compton heating term is calculated as follows. First, the average energy 
of photons in the fluid frame $\epsilon$ is calculated and the equivalent photon 
temperature $\theta_\gamma=\epsilon/3$ is defined. 
Noting that $\langle \gamma^2-1 \rangle =3\theta K_3(1/\theta)/K_2(1/\theta)$, 
we put 
\begin{equation}
\epsilon_\gamma=3\theta_\gamma \frac{K_3(1/\theta_\gamma)}{K_2(1/\theta_\gamma)}. 
\end{equation}
The radiation field is assumed to be given by the free streaming approximation 
in the optically thin regime. 
The photons are emitted from the photosphere and 
they are relativistically beamed with the Lorentz factor of the photosphere. 
Although we take account of effects of photons on the dynamics of electrons, 
we neglect the corresponding change of the photon field for simplicity. 
The equilibrium Lorentz factor of the radiation field is given by \citep{siko96}
\begin{equation}
\Gamma_{\rm eq}=
{1 \over \sqrt{2}} \left( 1+{1 \over \sqrt{1-\delta^2}} \right)^{1/2}, 
\end{equation}
where
\begin{equation}
\delta={2 T_\gamma^{01} \over T_\gamma^{00} + T_\gamma^{11}}.
\end{equation}

For ${F'^{0}}$, we also include annihilation cooling term as
\begin{equation}
({F'^{0}})_{\rm ann} = 2 m_{\rm e} \dot{n_e} \langle \gamma \rangle
\end{equation}
in addition to the Compton cooling and heating term.
We neglect other terms such as bremsstrahlung cooling, pair production 
from the photon field. Thus, our treatment is necessarily simplified 
to obtain detailed estimate of the final states of the flow, 
but we believe that the basic features of the flow are not much lost  
by these simplifications.

\section{Analytical solution}

In this section we describe analytic approximation, which help 
to understand the physical situation considered in this paper.

\subsection{Dynamics of Wien equilibrium pair plasma}

In the optically thick regime, adopting one fluid approximation 
and assuming that Wien equilibrium (eq.\ref{suumitudohi}) is maintained in the flow too, we have two integrals from 
energy and number conservation laws. 
The energy conservation is written in the one fluid approximation from eqs. (\ref{2-1}),(\ref{2-4}),(\ref{2-5}),(\ref{2-6}),(\ref{2-7})\&(\ref{suumitudohi})
\begin{equation}
4\pi r^2 m_{\rm e} n_\gamma 
\left\{ 4 \theta + 2f(\theta)(\theta+ \langle \gamma \rangle) \right\} 
\Gamma^2 \beta = \dot{E}, \hspace{.5cm}  
\label{divide1}
\end{equation}
where $\dot{E}$ is the total luminosity.
Since the sum of the number of photons and pairs is assumed to be conserved, 
we obtain from eqs. (\ref{2-9})\&(\ref{2-10})
\begin{equation}
4\pi r^2 n_\gamma \{ 1+2f(\theta) \} \Gamma \beta = \dot{N}, \hspace{5mm} 
\label{divide2}
\end{equation}
where $\dot{N}$ is the total number flux of photons and leptons.

Dividing eq.(\ref{divide1}) by eq.(\ref{divide2}), we obtain
\begin{equation}
\Gamma ={\dot{E} \over m_{\rm e} \dot{N}} 
{1 + 2f(\theta) \over 4\theta + 2f(\theta)(\theta+ \langle \gamma \rangle)}.
\end{equation}
In the limit of relativistic temperature, $f(\theta) \to 1$ and 
$ \langle \gamma \rangle \to 3\theta$,
then we have
\begin{equation}
\Gamma \to {\dot{E} \over m_{\rm e} \dot{N}} {1 \over 4\theta}.
\end{equation}
In the limit of non-relativistic temperature, $f(\theta) \to 0$ and 
$\langle \gamma \rangle \to 1$, we have the same expression
\begin{equation}
\Gamma \to {\dot{E} \over m_{\rm e} \dot{N}} {1 \over 4\theta}.
\end{equation}
It is to be noted that this expression is not an exact one at intermediate 
temperatures, in contrast to the case of complete thermal equilibrium(GW98). 
However, we use this as an approximate analytic solution in this section. 
This relation implies that the bulk Lorentz factor increases as 
the temperature decreases, i.e., that the plasma is accelerated thermally 
in compensation of the internal energy. 
The factor $\dot{E}/m_{\rm e} \dot{N}$ represents 
the mean energy of particles except for a numerical factor of order unity.  
Supposing that $\Gamma$ at the boundary is order of unity, the bulk Lorentz 
factor attains an order of 10, if the plasma remains optically thick during 
an order of magnitude drop in the temperature. The temperature at the boundary 
needs not necessarily be highly relativistic as long as it is high enough 
to produce pairs. But, it is important to determine the ratio of pairs 
to photons at various radii. The actual kinetic power of pairs at infinity 
depends on the amount of surviving pairs too, which is considered in the 
following subsections. 

The photon number density is also represented by the temperature as 
(for $\Gamma \gg 1$)
\begin{equation}
n_\gamma = {\dot{N}^2 \over \dot{E}} {m_{\rm e} \over 4\pi r^2}
           {4\theta + 2f(\theta)(\theta+ \langle \gamma \rangle ) 
\over  \left\{ 1+2f(\theta) \right\} ^2}.
\end{equation}
At the relativistic temperatures, it becomes
\begin{equation}
n_\gamma = {\dot{N}^2 \over \dot{E}} {m_{\rm e} \over 4\pi r^2}
            {4\theta \over 3},
\end{equation}
while at non-relativistic temperature, it is 
\begin{equation}
n_\gamma = {\dot{N}^2 \over \dot{E}} {m_{\rm e} \over 4\pi r^2}
              4\theta. 
\end{equation}
The difference of the factor 3 means that at non-relativistic temperatures 
all the pairs annihilate and are converted to photons.
Anyway, the total number flux of photons and electron pairs is conserved and 
total number density is approximated by 
\begin{equation}
n_\gamma +2 n_{\rm e} 
= {\dot{N}^2 \over \dot{E} } {m_{\rm e} \over 4 \pi r^2} 4 \theta. 
\label{num1}
\end{equation}

Under the approximations of  
$\Gamma =\dot{E}/4 m_{\rm e} \theta \dot{N}$ and the above expression 
for $n_\gamma+2n_{\rm e}$, the equation of motion (eq. \ref{2-3}) becomes 
\begin{equation}
-\frac{1}{\theta}\frac{d\theta}{dr}
+\frac{(1+2f(\theta))r^2}
           {\theta[4\theta+2f(\theta)(\theta+\langle\gamma\rangle)]}
\frac{d}{dr}(\frac{\theta^2}{r^2})=0.
\end{equation}
Thus, both for relativistic and non-relativistic temperatures, we obtain 
\begin{equation}
\theta \propto r^{-1}.
\end{equation}
Then, we obtain $\Gamma \propto r$ and $n_\gamma+2n_{\rm e} \propto r^{-3}$.
Thus, thermal acceleration works for this Wien equilibrium 
pair outflows. 

In order that optical thick condition is satisfied, 
the optical thickness at the boundary should be at least larger than unity. 
We suppose that the boundary temperature $\theta_0$ is at least 
mildly relativistic and the number densities of photons and electrons 
are comparable ($f(\theta_0)\simeq 1$). Thus, we have 
\begin{eqnarray}
n_{\gamma, 0}  \simeq  n_{\rm e, 0} & 
\simeq & {1 \over 3}(n_{\gamma, 0}+2n_{\rm e, 0}) \nonumber \\ 
 &=& {1 \over 3}{m_{\rm e} \over \pi r_0^2}
{\dot{N}^2 \over \dot{E}} \theta_0 \nonumber \\
 & \sim & {1 \over 48 \pi m_{\rm e}} 
{\dot{E} \over r_0^2 \theta_0 {\Gamma_0}^2}.
\label{num2}
\end{eqnarray}
From eq.(\ref{tau}), the optical thickness at the boundary is estimated as
\begin{equation}
\tau_0 \simeq {n_{{\rm e}, 0} \sigma_T r_0 \over 3\Gamma_0}
\simeq {1 \over 72}{r_g \over r_0}{m_{\rm p} \over m_{\rm e}}
{1 \over {\Gamma_0}^3 \theta_0}{\dot{E} \over L_{\rm Edd}},
\label{tau1}
\end{equation}
where the Eddington luminosity is defined by 
\begin{equation}
L_{\rm Edd} \equiv 2 \pi {m_{\rm p} r_{\rm g} \over \sigma_T}
\end{equation}
and $r_{\rm g}$ is the gravitational radius.
Thus, in order to realize $\tau_0 \gg 1$, the total luminosity 
should be at least an order of the Eddington luminosity for 
reasonable ranges of $\Gamma_0$ and $\theta_0$ at the boundary 
$r_0=2r_{\rm g}$. 

The optical thick solution obtained in this subsection is summarized as 
\begin{equation}
\Gamma=\Gamma_0 \frac{r}{r_0},
\end{equation}
\begin{equation}
\theta=\theta_0 \frac{r_0}{r}
\end{equation}
and
\begin{equation}
n_\gamma +2 n_{\rm e} = 3n_{\rm e,0} (\frac{r_0}{r})^3.
\end{equation}

\subsection{Photosphere and freeze-out radius}

The optical thickness decreases with radius and the flow 
eventually becomes optically thin. 
For relativistic temperatures, the optical thickness is given by 
\begin{equation}
\tau \simeq {n_{\rm e} \sigma_T r \over 3\Gamma} 
\simeq {n_{\rm e, 0} \sigma_T r_0 \over 3\Gamma_0} 
\left( {r \over r_0} \right)^{-3} 
\equiv \tau_0 \left( {r \over r_0} \right)^{-3}.
\end{equation}
Defining the radius of the photosphere $r_{\rm ph}$ as the radius where 
$\tau=1$, we obtain 
\begin{equation}
r_{\rm ph} \simeq \tau_0^{1/3} r_0.
\end{equation}
The temperature at the photosphere $\theta_{\rm ph}$ is 
\begin{equation}
\theta_{\rm ph} \simeq \theta_0 \tau_0^{-1/3}.
\end{equation}
If $\theta_{\rm ph}$ is smaller than unity, the real photospheric 
radius is smaller than $r_{\rm ph}$ given above because $n_{\rm e}$ 
becomes smaller than $n_\gamma$. 

On the other hand, equating annihilation time scale with dynamical 
time scale, we obtain the freeze-out radius $r_{\rm fr}$ as 
\begin{eqnarray}
 & & {n_{\rm e} \over (\dot{n_{\rm e}})_{\rm ann}} 
\simeq {r_{\rm fr} \over \Gamma} \nonumber \\
\Rightarrow & &
r_{\rm fr} \simeq \left( {9 \over 8}\tau_0 \right)^{1/3} r_0 
\simeq \left( {9 \over 8} \right)^{1/3} r_{\rm ph}. 
\end{eqnarray}
These two radii are almost the same. Although the above expression 
for $r_{\rm fr}$ is correct for $\theta_{\rm ph}>1$, the coincidence 
of the two radii itself is valid even if $\theta_{\rm ph}<1$. 
Thus, copious pairs exist 
in the optically thick regime because of the photon photon pair 
production counterbalancing the pair annihilation, 
unless the temperature becomes too low at $r_{\rm ph}$. 
In the optically thin regime, a large fraction of pairs can 
survive because the annihilation 
time scale becomes longer than the dynamical time scale. 

The behavior of the freeze-out depends on the 
temperature at the photosphere $\theta_{\rm ph}$. 
For $\theta_{\rm ph}<1$, the number density of pairs 
becomes small compared to the photon density but pairs can still carry 
a significant amount of kinetic power in the rest mass energy. 
For $\theta_{\rm ph}>1$, the number density of pairs is comparable to 
the photon density and the kinetic power of pairs accounts for 
more than half of the total power. Surviving kinetic power, however, 
is affected by interactions with radiation in the optically thin 
regime.  
We call the former and latter non-relativistic freeze-out and 
relativistic freeze-out, respectively.
The condition for relativistic freeze-out is given by 
\begin{equation}
\theta_{\rm ph} = \left( {r_{\rm ph} \over r_0} \right)^{-1} \theta_0 >1 
\ \Longleftrightarrow \
\theta_0 > {\tau_0}^{1/3}. 
\end{equation}
In Figure 1 are shown these regimes in the 
$\theta_0-\dot{E}$ plane, where we assume $r_0=2 r_g$ and  
$\Gamma_0=\sqrt{3/2}$. 
Thus, $\dot{E}/L_{\rm Edd}=0.144\tau_0 \theta_0$ holds.

\begin{figure}
\epsscale{1}
\plotone{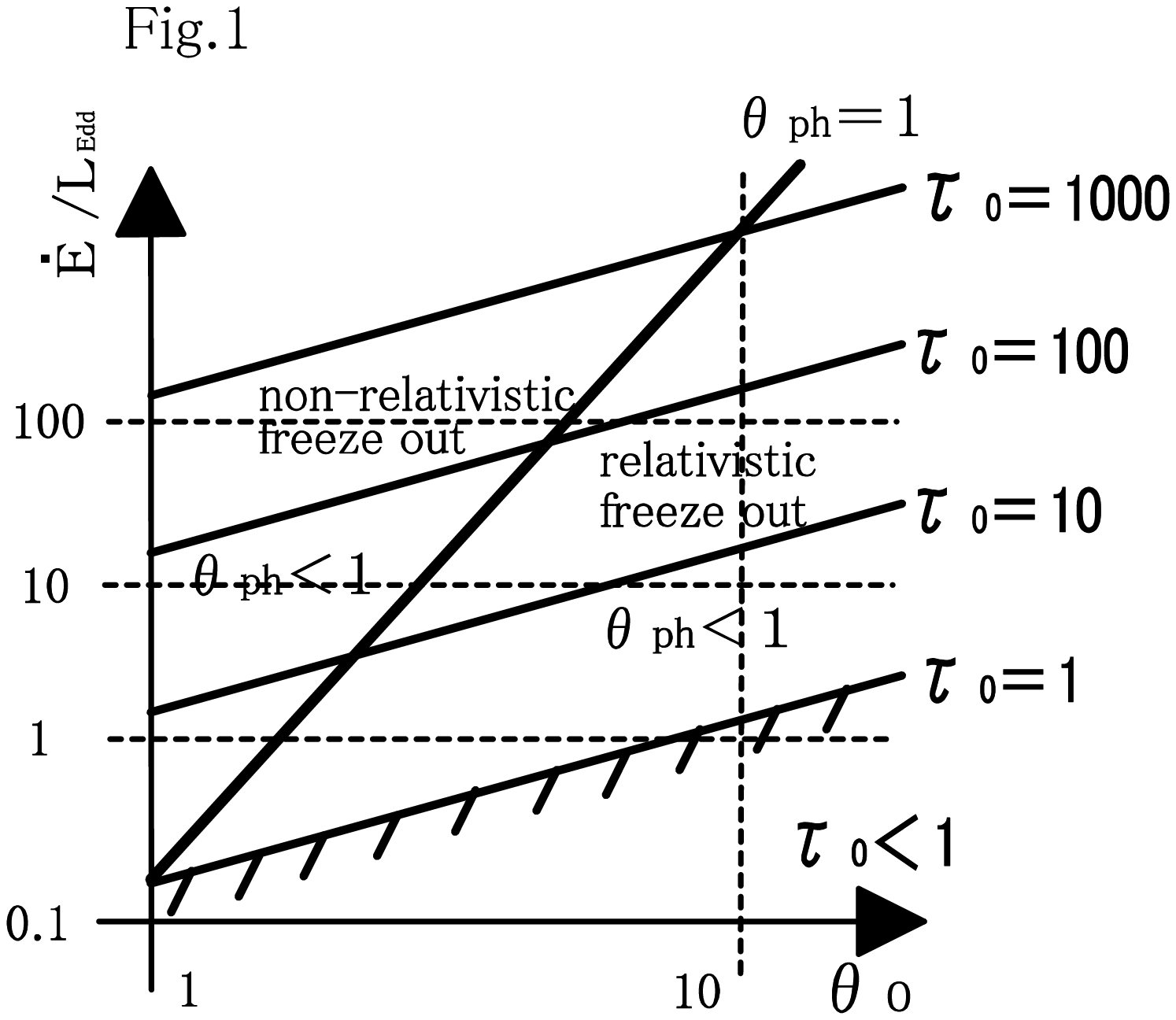}
\caption{Photospheric temperature in the $\dot{E}-\theta_0$ plane. 
The upper-left region of $\theta_{\rm ph}=1$ corresponds to 
the non-relativistic freeze-out with $\theta_{\rm ph}<1$, while 
the lower-right region does to the relativistic freeze-out with
$\theta_{\rm ph}>1$. The lines of $\tau_0=$const. are also depicted. }
\label{freeze}
\end{figure}

\placefigure{freeze}

\subsection{Relativistic freeze-out}

When $\theta_0>\tau_0^{1/3}$ is satisfied, 
the photospheric temperature is relativistic and  
we approximate $f(\theta_{\rm ph})=1$ and neglect pair annihilation 
in the optically thin regime.
Thus, the number flux of pairs $\dot{N}_{\rm e}$ is constant from 
the boundary to infinity; it is given by 
\begin{equation}
\dot{N}_{\rm e\pm , \infty} \simeq \dot{N}_{\rm e\pm , 0} 
\simeq {2 \over 3}\dot{N}
\end{equation}
The kinetic power of pairs at the photosphere is also estimated as 
\begin{equation}
\dot{E}_{\rm e \pm, ph} \simeq \frac{2}{3}\dot{E} .
\end{equation}
The bulk Lorentz factor at the photosphere is 
\begin{equation}
\Gamma_{\rm ph} \simeq \tau_0^{1/3} .
\end{equation}

After freeze-out, these pairs still interact with radiation in 
the optically thin regime. Since in the regime of relativistic temperature, 
the pair temperature and equivalent photon temperature seen in 
the comoving frame of pairs behave in the same way, the Compton 
cooling/heating effects may not be large and the internal energy 
is further converted to the bulk acceleration. If the internal energy at 
the boundary is converted to the bulk acceleration, we expect  
$\Gamma_\infty \sim 4\theta_0$.
After the pair temperature becomes non-relativistic, the pair temperature 
becomes lower than the equivalent photon temperature seen in the 
comoving frame of pairs. Then, we expect Compton heating rather than the 
Compton cooling there. In addition, since the photons are strongly beamed, 
some amount of radiative acceleration is also expected. 
Thus, we expect that interaction with photons transfers energy from 
photons to pairs, although the details should be investigated by 
numerical calculations.  

To summarize, we expect that the terminal state of the pair outflow 
for relativistic freeze-out 
is characterized by 
\begin{equation}
\Gamma_\infty \sim 4\theta_0
\end{equation}
and 
\begin{equation}
\dot{E}_{\rm e \pm,\infty} \sim \frac{2}{3}\dot{E}. 
\end{equation}
Thus, a Wien fireball turns out to be a very efficient mechanism 
for the production of relativistic outflows of electron-positron pairs 
at least for relativistic freeze-out.

\subsection{Non-relativistic freeze-out}

\placefigure{ann}
\begin{figure}
\epsscale{1}
\plotone{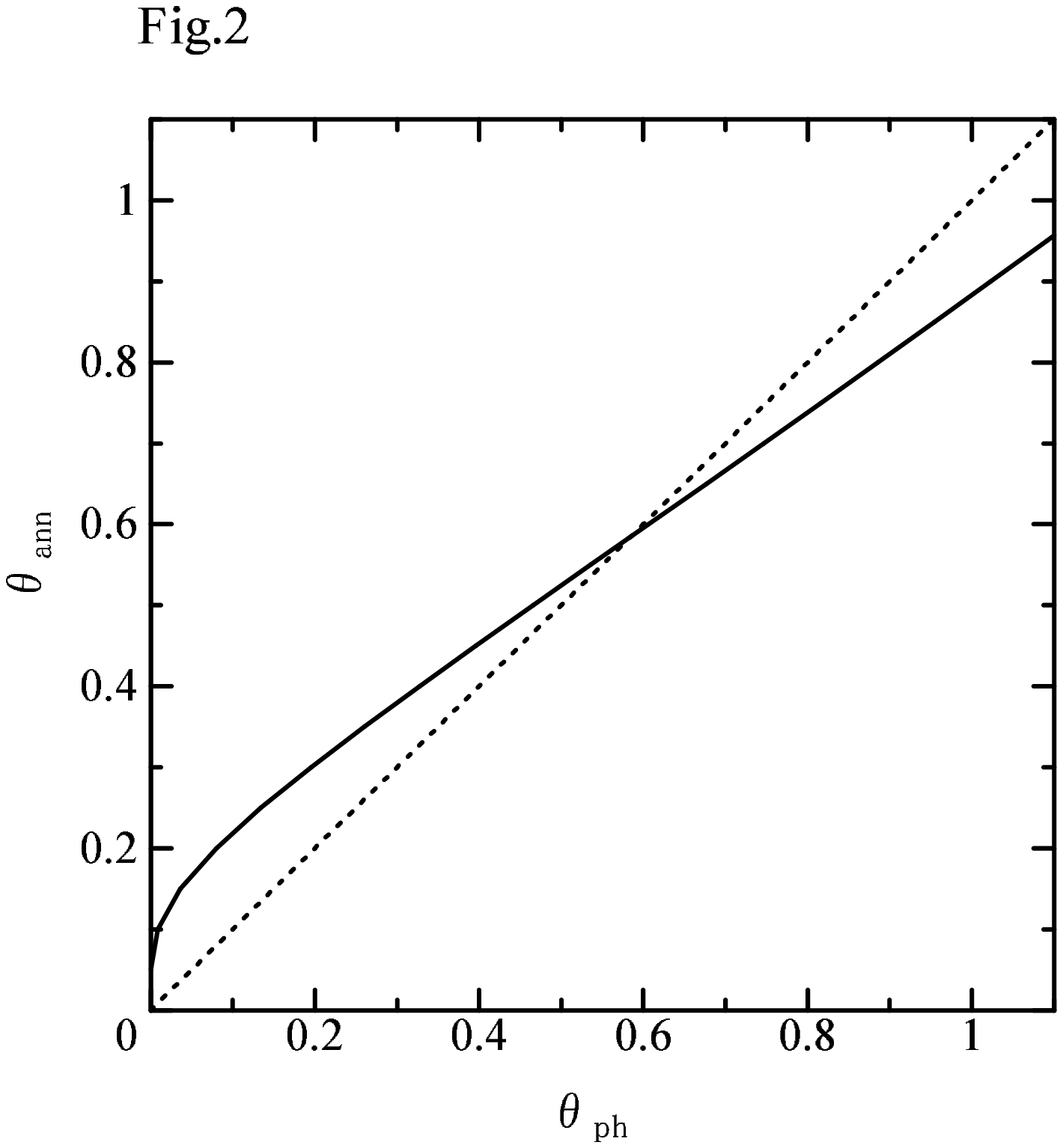}
\caption{Relation between $\theta_{\rm ann}$ and $\theta_{\rm ph}$. 
The solution to equation (\ref{ann}) is shown. 
The dotted line represents $\theta_{\rm ann}=\theta_{\rm ph}$. }
\label{ann}
\end{figure}

When $\theta_0<\tau_0^{1/3}$ is satisfied, 
the photospheric temperature $\theta_{\rm ph}=\theta_0 \tau_0^{-1/3}$ 
becomes non-relativistic. For $\theta \ll 1$, $f(\theta)$
behaves as \citep{sven84}
\begin{equation}
f(\theta) \approx \sqrt{\frac{\pi g(\theta)}{8\theta^3}}
\exp(-\frac{1}{\theta}),
\end{equation}
where $g(\theta)$ is approximated as 
\begin{equation}
g(\theta) \approx 1+3.7622\theta+5.1054\theta^2
+\frac{8}{\pi}\theta^3. 
\end{equation}
Thus, the equilibrium number density of electrons at $\theta_{\rm ph}$ 
decreases with decreasing $\theta_{\rm ph}$.
In this rough approximation, the kinetic power of pairs at the 
photosphere is given by 
\begin{equation}
\dot{E}_{\rm e \pm, ph} 
\simeq \dot{E}\frac{f(\theta_{\rm ph})}{2\theta_{\rm ph}}
\sim \dot{E}\sqrt{\frac{\pi g(\theta_{\rm ph})}{32\theta_{\rm ph}^5}}
\exp(-\frac{1}{\theta_{\rm ph}}).
\label{kin1}
\end{equation}
It is seen that the kinetic power of pairs can still be a fair fraction 
of the total luminosity when the photospheric temperature is mildly 
relativistic. For example, $\dot{E}_{\rm e\pm,ph}=0.16\dot{E}$ even for 
$\theta_{\rm ph}=0.2$. 

In reality, the pairs may not keep a Wien equilibrium since the 
pair creation becomes ineffective at non-relativistic temperatures. 
Thus, the pair annihilation process proceeds in a  
non-equilibrium way and the actual pair density is higher than 
the equilibrium value.  
This process can be treated in the same way as GW98. 
Let us first assume that the deviation from equilibrium is small and 
expand the electron number density as 
\begin{equation} 
\delta n_{\rm e} \equiv n_{\rm e} - n_{\rm e, eq} \ll n_{\rm e, eq}
\end{equation}
and 
\begin{equation}
n_{\rm e}^2 - {n_{\rm e, eq}}^2 \approx 2 n_{\rm e, eq} \delta n_{\rm e}.
\end{equation}
Since the electron number density follows the equation 
\begin{equation}
\dot{n}_{\rm e} 
=-\langle \sigma_{\rm ann} v \rangle  
\left[ {n_{\rm e}}^2 - { n_{\rm e, eq}}^2 \right] 
= {1 \over r^2} {d \over dr} \left[ n_{\rm e} r^2 \Gamma \beta \right],
\end{equation}
we obtain 
\begin{equation}
\delta n_{\rm e} 
\approx - {1 \over 2 n_{\rm e, eq}} 
{1 \over \langle \sigma_{\rm ann} v \rangle} 
{1 \over r^2} {d \over dr} \left[ n_{\rm e, eq}r^2 \Gamma \beta \right].
\end{equation}
Since $\Gamma \propto r$ as long as the equilibrium is maintained,
we obtain
\begin{equation}
\delta n_{\rm e} 
\approx - {\Gamma \over 2r \langle \sigma_{\rm ann} v \rangle } 
\left[ 3 + {d \ln n_{\rm e, eq} \over d \ln r} \right] 
\label{deltan}
\end{equation}
to the leading order.
From equation (\ref{num1}), 
equilibrium number density of electrons is estimated as
\begin{eqnarray}
n_{\rm e, eq} 
&=& {\theta m_{\rm e}\dot{N}^2 \over  \pi r^2 \dot{E}} 
{f(\theta) \over 2f(\theta)+1} \nonumber \\
& \sim & { m_{\rm e}\dot{N}^2 \over  \pi r^2 \dot{E}}
 \sqrt{{\pi g(\theta) \over 8 \theta}} \exp(-\frac{1}{\theta}).
\end{eqnarray}

Then, equation (\ref{deltan}) can be approximated as
\begin{equation}
\delta n_{\rm e} 
\sim - {\Gamma \over 2r \langle \sigma_{\rm ann} v \rangle} 
\left[ {3 \over 2} - {1 \over \theta} \right] 
\sim {\Gamma \over 2r \langle \sigma_{\rm ann}v \rangle} 
{1 \over \theta}
\end{equation}
for $\theta \ll 1$.

We regard that the freeze-out occurs when 
$\delta n_{\rm e}=n_{\rm e, eq}$ is satisfied. 
We define the annihilation radius $r_{\rm ann}$ as the radius 
where $\delta n_{\rm e}=n_{\rm e, eq}$ is satisfied 
to discriminate it from $r_{\rm fr}$ and $r_{\rm ph}$. 
The temperature at $r_{\rm ann}$ is denoted by $\theta_{\rm ann}$. 
The equation to determine $r_{\rm ann}$ and $\theta_{\rm ann}$ 
is given by 
\begin{equation}
{m_{\rm e}\dot{N}^2 \over \pi r_{\rm ann}^2 \dot{E}}
\sqrt{{\pi \over 8 \theta_{\rm ann}}} \exp(-\frac{1}{\theta_{\rm ann}})
={\Gamma_{\rm ann} \over 2r_{\rm ann} \langle \sigma_{\rm ann}v \rangle} 
{1 \over \theta_{\rm ann}}.
\end{equation}
Recalling that $\Gamma_{\rm ann} \simeq \Gamma_0(r_{\rm ann}/r_0)$ and 
$r_{\rm ann}=r_0(\theta_0/\theta_{\rm ann})$ and using 
equations (\ref{num2}) and (\ref{tau1}), we obtain the equation to determine 
$\theta_{\rm ann}$ as 
\begin{equation}
\theta_{\rm ann}^{5/2}\exp(-\frac{1}{\theta_{\rm ann}})
=\frac{16}{27 \sqrt{2\pi}} \frac{\theta_0^3}{\tau_0}
=\frac{16}{27\sqrt{2\pi}}\theta_{\rm ph}^3.
\label{ann}
\end{equation}
In Figure 2 the relation between $\theta_{\rm ann}$ and $\theta_{\rm ph}$ 
is shown and we see that the former is slightly higher than the latter 
for $\theta_{\rm ph} < 0.6$. 
The kinetic power of pairs estimated at the annihilation radius is 
given by 
\begin{equation}
\dot{E}_{\rm e \pm, ann} 
\simeq \dot{E}\frac{f(\theta_{\rm ann})}{\theta_{\rm ann}}
\sim \frac{4}{27}\dot{E}(\frac{\theta_{\rm ph}}{\theta_{\rm ann}})^3
\frac{1}{\theta_{\rm ann}^2}.
\end{equation}
This simple estimate gives a somewhat higher value than that by 
equation (\ref{kin1}). Thus, the kinetic power of pairs is shown to be 
still fairly high and is comparable to the radiative power even 
for non-relativistic freeze-out. As for the Lorentz factor at the photosphere,  
we expect that 
\begin{equation}
\Gamma_{\rm ph} \sim \frac{\theta_0}{\theta_{\rm ph}}.
\end{equation}
The terminal Lorentz factor may be higher by a factor of a few 
owing to the additional thermal and radiative acceleration 
in the optically thin regime.

\section{Numerical Results}

In this section, we present the numerical results which are compared with 
analytic predictions made in the previous section. 
Numerical calculations are performed for a wide range of values of 
the total luminosity $\dot{E}$ and the boundary temperature $\theta_0$ 
at $r=r_0=2r_{\rm g}$. 
We fix $\Gamma_0=\sqrt{3/2}$; other boundary values $\tau_0$, 
$n_{\gamma,0}$ and $n_{\rm e,0}$ and the total number flux $\dot{N}$ 
are easily obtained through several relations described in the previous 
section. 
We solve eqs. (\ref{2-1}),(\ref{2-3}),(\ref{2-9})\&(\ref{2-10})
for optically thick regime and eqs. (\ref{2-14}),(\ref{2-15})\&(\ref{2-16}) for optically thin regime.
The radiation field in the optically thin regime is 
calculated in the free streaming approximation as was 
described in section 2.
The photospheric radius which marks the boundary between these two regimes is determined iteratively.

As a typical example of non-relativistic freeze-out, we show in 
Figure 3 the numerical results for 
$\dot{E}/L_{\rm Edd}=24.9$ and $\theta_0=2$ with $\tau_0=82.4$.
Figure 3(a) shows the behavior of the bulk velocity $\Gamma\beta$ 
and temperature $\theta$. 
It is seen that the photosphere is located at $r_{\rm ph}=7.45r_{\rm g}$ 
and that the temperature decreases rapidly with radius and becomes 
$\theta_{\rm ph}=0.41$ at the photosphere. The bulk Lorentz factor 
at the photosphere turns out to be $\Gamma_{\rm ph}\beta_{\rm ph}=5.06$. 
These numerical values reasonably agree with the analytic predictions 
of $r_{\rm ph}=8.7 r_g$, $\theta_{\rm ph}=0.46$ and $\Gamma_{\rm ph}=5.3$. 
In particular, it is seen that a factor of 5 decrease in temperature 
matches very well a factor of 5 increase of the bulk Lorentz factor, which 
confirms the basic feature of the thermal acceleration of the pair plasma 
in the optically thick regime. 
The differences between analytic and numerical results may be due to  
some simplifications made in the analytic estimates such as $\Gamma \gg 1$ 
and pair annihilation process. 

It is particularly important that pair outflow is further  
accelerated outside the photosphere. The numerical result shows that 
the terminal Lorentz factor attains $\Gamma_\infty=15.9$, more than 
three times that at the photosphere. Most of this factor is accounted 
for by the internal energy of pairs at the photosphere with an additional  
effect of radiative acceleration. It should be noted that since even at 
$\theta_{\rm ph}=0.41$, $\theta+\langle\gamma\rangle=2.3$, 
pure thermal acceleration can attain $\Gamma_\infty\sim 12$. 
Moreover, beamed photons from the photosphere 
heat the pairs by Compton interaction since 
the effective temperature of photons is larger than electron 
temperature outside the photosphere as is seen in this figure. 
So the thermal acceleration works efficiently while radiative 
acceleration also makes a small contribution to bulk acceleration 
at larger distances. This additional bulk acceleration by a 
factor of 3 is an important element in comparison with observations. 

Figure 3(b) shows the relative number fraction of pairs and photons 
as a function of the radial coordinate. 
As is seen, the pair fraction decreases steadily according to the decrease 
of temperature and annihilation continues to $2r_{\rm ph}$.  
Thus, the analytic estimate of $r_{\rm fr}$ and $r_{\rm ann}$ may be 
uncertain within a factor of 2. 
But, the degree of the decrease is modest and about 20\% of pairs 
annihilate outside the photosphere. To sum up, 
about 60\% of the initial pairs survive in this case. 
This confirms our prediction that the pair annihilation problem can be 
avoided when we consider an initial Wien equilibrium state. 

\begin{figure}
\epsscale{1}
\plotone{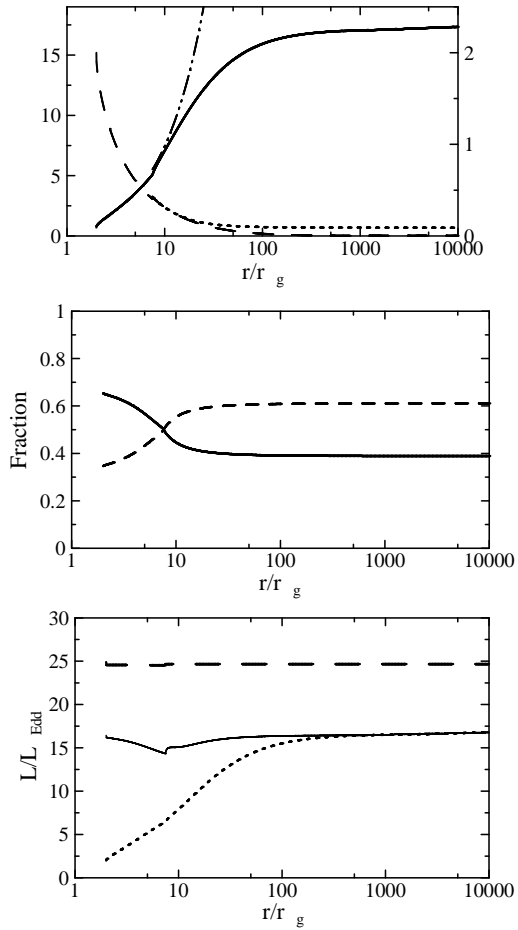}
\caption{
(a) Behavior of velocity and temperature for 
$\dot{E}/L_{\rm Edd}=24.9$, 
$\theta_0=2$ with $\tau_0=82.4$. The solid line denotes 
the bulk Lorentz factor of the pairs $\Gamma\beta$,
and the dash-dotted line denotes the equilibrium Lorentz factor 
of radiation field $\Gamma_{\rm eq}\beta_{\rm eq}$.
The dashed line denotes the temperature of pairs, and dotted line 
denotes the equivalent temperature of radiation outside the photosphere.
(b) Fraction of particle and photon numbers for 
$\dot{E}/L_{\rm Edd}=24.9$, $\theta_0=2$ with $\tau_0=82.4$.
The solid and dashed lines denote the number fractions of pairs and 
photons, respectively.
(c) Luminosities of various components for 
$\dot{E}/L_{\rm Edd}=24.9$, $\theta_0=2$ with $\tau_0=82.4$.
The dashed line denotes the total luminosity of pairs 
and radiation. The solid line denotes the kinetic power of pairs 
while the dotted line denotes the power carried in a form of the rest mass of 
pairs. Thus, the interval between the dashed and solid lines denotes 
the luminosity of radiation and that between the solid 
and dotted lines denotes the power carried in a form of the thermal energy 
of pairs.
}
\label{one-example}
\end{figure}

\placefigure{one-example}

Figure 3(c) shows the luminosity of pairs and radiation as a function of the 
radial coordinate. 
The total luminosity turns out to remain constant as a function of radius as 
it should be. Inside the photosphere, the kinetic power of pairs 
slightly decreases with radius, which is because pairs are converted to 
photons as the temperature decreases keeping a Wien equilibrium. 
The kinetic power of pairs is kept almost constant outside the photosphere 
as it should be for bulk acceleration by their own thermal energy.   
A small change is due to the interaction with photons. 
But, it is seen that the energy gained from 
Compton interaction almost cancels out with a small but finite amount 
of annihilation cooling. The terminal kinetic power of pairs turns out to be 
61\% of the total luminosity, while remaining 39\% is carried away by 
photons emitted from the photosphere. 
Since the matter at the photosphere is moving at $\Gamma=5$,  
distant observers see the radiation peaked at an energy of 
$\Gamma\theta_{\rm ph}\sim 2$, i.e., around 1MeV.
Thus, this model predicts that strong MeV emission should accompany 
the relativistic jet formation.  

\begin{figure}
\epsscale{1}
\plotone{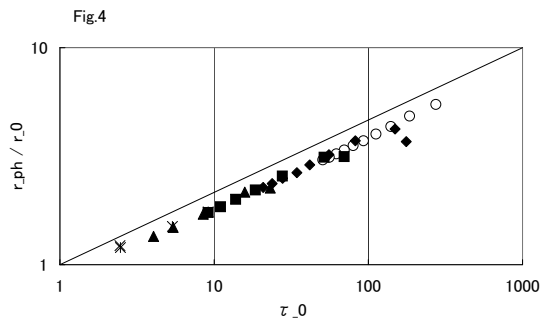}
\caption{
Relation between photospheric radius 
$r_{\rm ph}$ and the initial optical thickness $\tau_0$.
The solid line represents the analytic estimate 
$r_{\rm ph} / r_0 = {\tau_0}^{1/3}$. 
Open circles, solid diamonds, rectangles, triangles, crosses and 
stars refer to $\dot{E}/L_{\rm Edd}=83.1, 24.9, 8.31, 2.49, 0.831$ 
and $0.249$, respectively. 
}
\label{photosphere}
\end{figure}

\placefigure{photosphere}

\begin{figure}
\epsscale{1}
\plotone{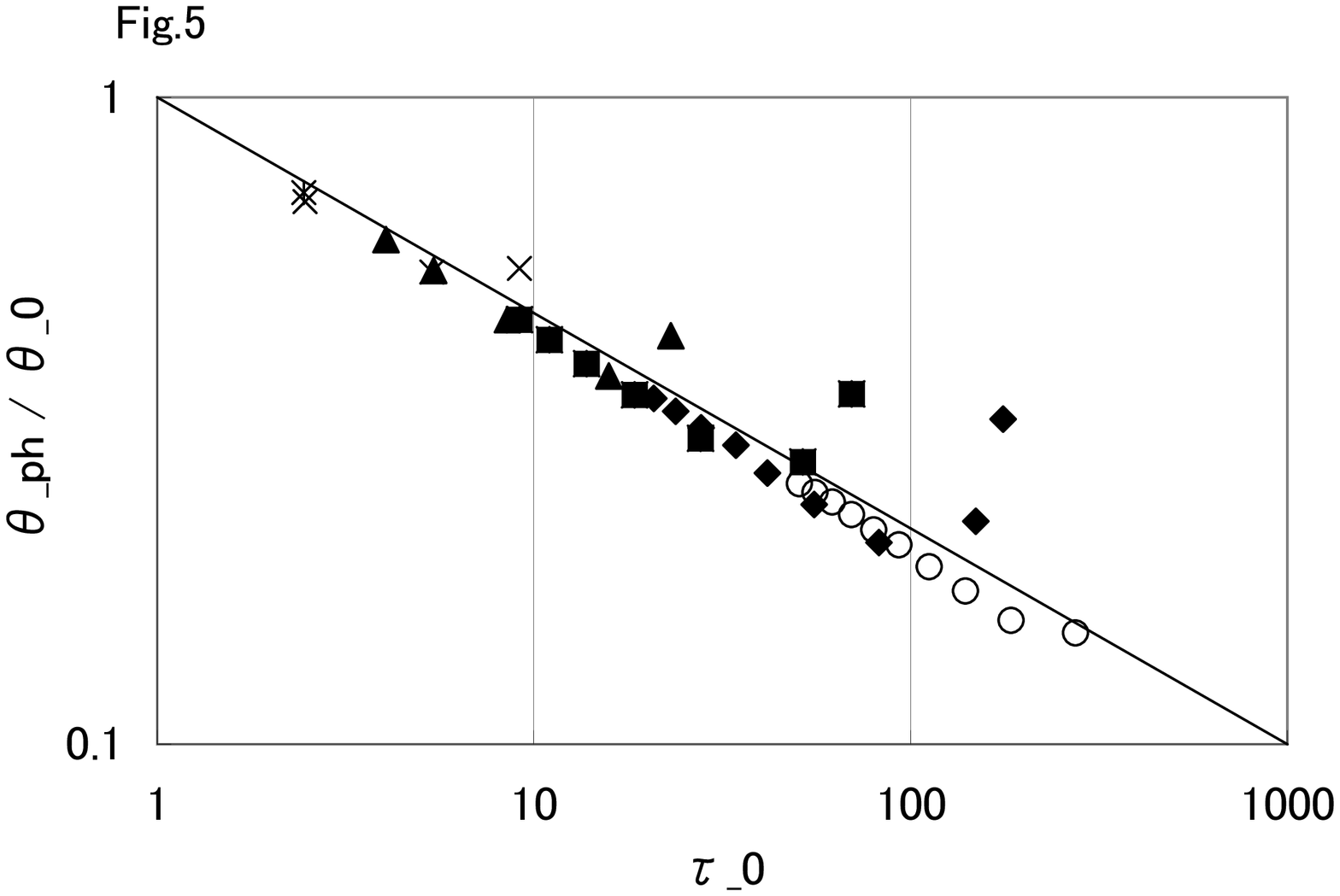}
\caption{
Relation between photospheric temperature 
$\theta_{\rm ph}$ and the initial optical thickness $\tau_0$.
The solid line represents the analytic estimate 
$\theta_{\rm ph} / \theta_0 = {\tau_0}^{-1/3}$. 
Open circles, solid diamonds, rectangles, triangles, crosses and 
stars refer to $\dot{E}/L_{\rm Edd}=83.1, 24.9, 8.31, 2.49, 0.831$ 
and $0.249$, respectively. 
}
\label{photosphere2}
\end{figure}

\placefigure{photosphere2}

\begin{figure}
\epsscale{1}
\plotone{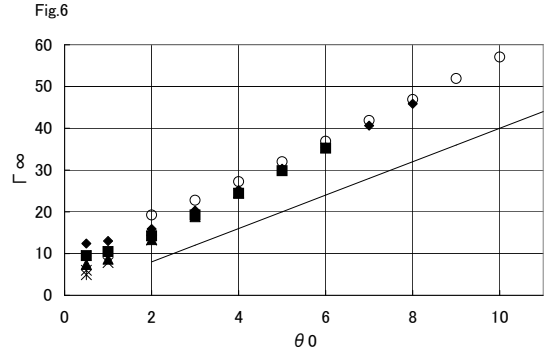}
\caption{
Terminal Lorentz factor $\Gamma_{\infty}$ 
for various boundary values. 
The abscissa is the temperature at the boundary $\theta_0$.
The symbols are the same as Fig. 4. 
The solid line represents a naive expectation of $\Gamma_{\infty} = 4 \theta_0$. 
}
\label{ginf}
\end{figure}

\placefigure{ginf}

We have also calculated for various boundary values and the results are 
tabulated in Table 1. 
As is seen, the terminal Lorentz factor proves to be $8 \sim 60$ 
and the terminal kinetic power of pairs accounts for $ 50 \sim 80 \%$ 
of the total luminosity for $1< \theta_0<5$ and 
$0.8 < \dot{E}/L_{\rm Edd} < 80$. 
Thus, the Wien fireball turns out to be quite an efficient mechanism for 
the production and bulk acceleration of relativistic jets. 
As for the dependence of the physical quantities at the photosphere, 
they reasonably agree with the analytic predictions of  
$r_{\rm ph} \propto \tau_0^{1/3} \propto \dot{E}^{1/3}\theta_0^{-1/3}$, 
$\theta_{\rm ph} \propto \dot{E}^{-1/3}\theta_0^{4/3}$ and 
$\Gamma_{\rm ph} \propto \dot{E}^{1/3}\theta_0^{-1/3}$. 
Figure 4 shows $r_{\rm ph}$ as a function of $\tau_0$ and 
figure 5 does $\theta_{\rm ph}$ as a function of $\tau_0$ and then, 
we see that
the analytic prediction agrees with numerical results, although 
the analytic prediction is slightly larger than the numerical result. 

Since the enthalpy of pairs at the photosphere is large, 
pairs are thermally accelerated outside the photosphere. 
The terminal Lorentz factor is thus basically determined by 
the boundary temperature for relativistic freeze-out. 
For non-relativistic freeze-out, additional effect by 
radiation works to further increase the terminal Lorentz factor. 
This behavior is clearly seen in Figure 6. 
The terminal Lorentz factor
is above the naive estimate $4\theta$ to a varying degree  
according to a varying degree of importance of additional 
thermal and radiative acceleration. It is to be noted that 
fairly large Lorentz factor can be obtained even for the initial 
temperature as low as 0.5. 

Finally, in Figure 7 is shown the efficiency of the outflow production, 
i.e., the ratio of the terminal kinetic power of pairs to 
the total luminosity. It is basically determined by the photospheric 
temperature. For relativistic freeze-out, the efficiency is about 
10\% larger than the canonical value of 2/3. This excess is due to 
additional radiative acceleration. For non-relativistic freeze-out, 
the efficiency becomes small because pair density decreases owing to 
pair annihilation. At $\theta_{\rm ph}=0.2$ the efficiency is about 30\% 
and it will be very small for lower $\theta_{\rm ph}$.  

Summarizing these results, we have found that electron-positron pairs 
can be thermally accelerated up to the bulk Lorentz factor of more than 10 
and that their kinetic power becomes comparable to the total power, 
provided that Wien equilibrium states of pure electron-positron pairs 
are prepared at the inner boundary 
with a temperature of a few times electron mass and the optical 
thickness to the scattering of more than around 5. 
The corresponding total luminosity exceeds the Eddington luminosity 
by a factor of more than unity in the spherical symmetry. 
If we assume that the outflow is conical and that the opening angle of 
the outflow covers $\Omega$ steradian, the total power is smaller by a factor 
of $4\pi/\Omega$. Since $\Omega$ is typically as small as $10^{-2}$, 
the required total luminosity can be as small as 1\% of the 
Eddington luminosity. 
The mechanisms of the realization of pure electron-positron pairs 
and initial collimation of the outflow remain to be a theoretical challenge. 
However, we believe that this is the most successful specific mechanism of 
the bulk acceleration of relativistic outflow up to the Lorentz factor of 10 with 
high efficiency. This Wien fireball model predicts that 
radiation from the photosphere should appear as a strong emission around  
a few MeV with a comparable luminosity to the jet kinetic power.

\begin{figure}
\epsscale{1}
\plotone{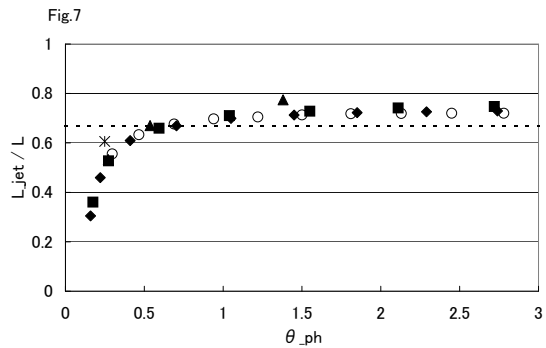}
\caption{
Efficiency of the outflow production; i.e., the ratio of 
the terminal kinetic power of pairs to the total luminosity.
The abscissa is the photospheric temperature. 
The symbols are the same as in Fig. 5.
The dotted line refers to the efficiency of 2/3.
}
\label{efficiency}
\end{figure}

\placefigure{efficiency}

\section{Discussion}

Finally, we discuss several issues related to the Wien fireball model. 
The first problem is the realization of the initial state. 
Most probably such a state is prepared in relation to the hot accretion disks 
or hot accretion disk corona. Previous studies on pair concentration in 
accretion disks have shown that the pair concentration in the disk is not 
so large under the steady state condition and that the steady state solution 
does not exist when the luminosity is higher than about a few percent 
of the Eddington luminosity \citep{KT88}. This is due to the inefficiency of the 
pair annihilation and suggests that pair runaway occurs for such 
high luminosity. Instead of the pair equilibrium condition, if we allow 
free escape of pairs from the disk, \cite{yama99} have shown that 
much higher pair concentration is obtained and that outflow luminosity of 
pairs can be very large based on a  
simple analysis. This seems to be a promising way to produce 
the initial state of the present outflow model, since \cite{yama99} predicts 
the optical thickness of pairs is around a few and the temperature of a few. 
Although pure pair escape was assumed in \cite{yama99},  
in reality baryon load should be carefully examined. 
If the baryon density is more than 1\% of the pair density, 
the terminal Lorentz factor will be significantly decreased. 
This is a common problem with the thermal fireball for GRBs. 

As for the photon field, we assumed that Wien equilibrium photons 
are prepared at the boundary and subsequent photon production process 
is neglected except for pair annihilation. If the initial optical 
thickness is so large, this seems to be a fair assumption but 
for optically thickness of a few to 10, this is not true and 
photon spectrum will not be a Wien one but near the Comptonized bremsstrahlung. 
Moreover, in order to have a large pair concentration, the amount of soft photons 
must be small because they will decrease the electron temperature and 
pair production. In any case, radiative transfer including such as 
bremsstrahlung and Compton scattering should be examined in a future work. This 
is also important to predict the accompanying photon spectrum with jet 
production. It is to be noted that whole process should be treated dynamically, 
since the pair escape is the most critical assumption for this mechanism. 

Our calculation indicates that super Eddington luminosity is needed for 
an relativistic jet production provided that the flow is spherically symmetric. 
As was mentioned in the previous section, if some collimation mechanisms exist 
near the boundary, the required 
luminosity will be smaller and the real luminosity can be 
sub-Eddington. 

For many years, radiative bulk acceleration mechanism has been regarded  
ineffective in producing relativistic jets. But most of them treat optically 
thin case \citep{phin82, IT96}. 
Since the equilibrium Lorentz factor of photons emitted from typical 
accretion disks is not large enough near the accretion disk, 
radiative force acts as  
radiation drag rather than as acceleration. So the attainable Lorentz 
factor of bulk flow is at most about 3.
This is because radiation field is not collimated enough to the jet direction. 
Further, for optically thin plasma, only a small fraction of the radiation 
power is converted to plasma kinetic power. Thus, the optically thin 
radiative acceleration is ineffective in general. In contrast, our present 
model assumes that the pair plasma starts to flow out with optically thick 
condition. This means that photons move at the same speed with pairs 
since they are strongly coupled and that they can be thermally accelerated. 
Although the pair outflow becomes optically thin outside the photosphere, 
radiative acceleration acts effectively because radiation has been 
collimated at the photosphere and the equilibrium Lorentz factor becomes 
large enough. 
Of course, the radiation drag from ambient photons will 
somewhat decrease the bulk velocity of the jets, but the effect 
will be small since the 
beamed radiation field dominates over ambient photons for 
typical cases. 

Another argument against pair jets has been the annihilation problem 
\citep{CF93,BL95}. 
When a large amount of cold pairs exist in a compact region, they will 
inevitably annihilate before they escape. 
Since in our model pairs are hot and accompanied with high energy photons, 
pair production works as well. 
Inside the photosphere, pair annihilation 
is balanced with pair production and outside the photosphere 
annihilation time scale becomes longer than the dynamical time scale. 
Thus, the most pairs can survive to infinity and can avoid the annihilation  
problem with pair jets.

This work is supported in part by the Grant-in-Aid for Scientific Research 
of the Ministry of Education and Science No.11640236 and by 
Research Fellowships of the Japan Society for the Promotion of Science.




\clearpage

\begin{table}

\begin{center}
Table 1
\begin{tabular}{|rrr|rrr|rrr|}
\hline

$\dot{E} / L_{\rm Edd}$ & $\theta_0$ & $\tau_0$ & $r_{\rm ph}$ & 
$\Gamma_{\rm ph} \beta_{\rm ph}$ & $\theta_{\rm ph}$ & 
$\Gamma_\infty \beta_\infty$ & $L_{\rm jet}$ & $L_{\rm jet} / \dot{E}$ \\

\hline
\hline

0.249 & 0.5 & 2.45 & 2.45 & 1.30 & 0.356 & 4.95 & 0.151 & 0.606 \\

\hline

0.831 & 0.5 & 9.15 & 3.50 & 1.94 & 0.272 & 6.00 & 0.442 & 0.532 \\
0.831 & 1   & 5.36 & 3.00 & 1.87 & 0.537 & 7.84 & 0.557 & 0.670 \\
0.831 & 2   & 2.47 & 2.40 & 2.30 & 1.38  & 13.2 & 0.643 & 0.774 \\

\hline

2.49 & 0.5 & 23.1 & 4.51 & 2.52 & 0.214 & 7.40 & 1.10 & 0.443 \\
2.49 &   1 & 15.8 & 4.32 & 2.73 & 0.371 & 8.67 & 1.49 & 0.600 \\
2.49 &   2 & 8.50 & 3.42 & 2.38 & 0.906 & 13.4 & 1.78 & 0.716 \\
2.49 &   3 & 5.42 & 2.97 & 2.01 & 1.62  & 18.9 & 1.88 & 0.756 \\
2.49 &   4 & 4.05 & 2.70 & 1.78 & 2.41  & 24.5 & 1.96 & 0.786 \\

\hline

8.31 & 0.5 & 69.7 & 6.30 & 3.42 & 0.174 & 9.51 & 3.00 & 0.361 \\
8.31 &   1 & 51.7 & 6.26 & 3.77 & 0.273 & 10.5 & 4.38 & 0.527 \\
8.31 &   2 & 27.7 & 5.12 & 3.62 & 0.594 & 14.2 & 5.48 & 0.659 \\
8.31 &   3 & 18.5 & 4.42 & 3.24 & 1.04  & 19.2 & 5.90 & 0.710 \\
8.31 &   4 & 13.8 & 4.00 & 2.93 & 1.55  & 24.5 & 6.05 & 0.728 \\
8.31 &   5 & 11.0 & 3.70 & 2.70 & 2.11  & 29.8 & 6.16 & 0.741 \\
8.31 &   6 & 9.15 & 3.48 & 2.50 & 2.72  & 35.2 & 6.21 & 0.747 \\

\hline

24.9 & 0.5 & 176  & 7.38 & 4.09 & 0.159 & 12.4 & 7.59 & 0.305 \\
24.9 &   1 & 149  & 8.44 & 4.78 & 0.221 & 13.0 & 11.5 & 0.460 \\
24.9 &   2 & 82.4 & 7.45 & 5.06 & 0.410 & 15.9 & 15.2 & 0.609 \\
24.9 &   3 & 55.5 & 6.43 & 4.71 & 0.704 & 20.3 & 16.7 & 0.669 \\
24.9 &   4 & 41.7 & 5.77 & 4.35 & 1.05  & 25.2 & 17.4 & 0.699 \\
24.9 &   5 & 34.4 & 5.31 & 4.02 & 1.45  & 30.3 & 17.8 & 0.713 \\
24.9 &   6 & 27.8 & 5.00 & 3.79 & 1.85  & 35.4 & 18.0 & 0.723 \\
24.9 &   7 & 23.8 & 4.73 & 3.58 & 2.29  & 40.6 & 18.1 & 0.726 \\
24.9 &   8 & 20.8 & 4.54 & 3.43 & 2.74  & 45.9 & 18.2 & 0.729 \\

\hline

83.1 &  2 & 274  & 11.0 & 6.83 & 0.297 & 19.2 & 46.1 & 0.555 \\
83.1 &  3 & 185  & 9.68 & 6.79 & 0.466 & 22.8 & 52.6 & 0.633 \\
83.1 &  4 & 140  & 8.67 & 6.43 & 0.690 & 27.2 & 56.2 & 0.676 \\
83.1 &  5 & 112  & 8.00 & 6.11 & 0.940 & 32.0 & 57.9 & 0.697 \\
83.1 &  6 & 93.1 & 7.45 & 5.78 & 1.22  & 36.9 & 58.6 & 0.705 \\
83.1 &  7 & 79.8 & 7.08 & 5.48 & 1.50  & 41.9 & 59.3 & 0.713 \\
83.1 &  8 & 69.8 & 6.75 & 5.23 & 1.81  & 46.9 & 59.7 & 0.718 \\
83.1 &  9 & 62.0 & 6.48 & 5.05 & 2.13  & 51.9 & 59.8 & 0.719 \\
83.1 & 10 & 55.8 & 6.25 & 4.86 & 2.45  & 57.1 & 59.8 & 0.720 \\
83.1 & 11 & 50.7 & 6.08 & 4.72 & 2.78  & 62.1 & 59.8 & 0.720 \\

\hline

\end{tabular}

\caption{Numerical results for various boundary conditions. 
First, second and third columns refer to the total luminosity, 
boundary temperature and boundary optical thickness. The 4-th through 
6-th columns show the numerical results for the quantities at the photosphere, 
while 7-th through 9-th columns show the numerical results for 
those at infinity.}

\end{center}

\end{table}

\end{document}